\documentclass{mem}
\usepackage{natbib}
\usepackage{txfonts}
\usepackage{balance}
\usepackage{graphicx}
\usepackage[a4paper]{hyperref}
\idline{00}{000}
\begin{document}

\title{The \textit{Swift}/XRT Catalogue of GRBs}
\author{Elena Zaninoni\inst{1,2}\footnote{E-mail: elena.zaninoni@brera.inaf.it},
   Raffaella Margutti\inst{1,3},
   Maria Grazia Bernardini\inst{1,4},
   Guido~Chincarini\inst{1,5},
  et al.
  }
\institute{
 INAF Osservatorio Astronomico di Brera, via Bianchi 46, Merate 23807, Italy
  \and Universit\`a di Padova, Dip. Astronomia, v. dell' Osservatorio 31, Padova 35122, Italy
  \and Harvard-Smithsonian Center for Astrophysics, 60 Garden Street, Cambridge, MA02138
  \and ICRANet, p.le della Repubblica 10, Pescara 65100, Italy
  \and Universit\`a Milano Bicocca, Dip. Fisica G. Occhialini, P.zza della Scienza 3, Milano 20126, Italy\\
}
\authorrunning{E. Zaninoni}
\titlerunning{XRT Catalogue}
\onecolumn
\abstract{We present the preliminary analysis of the GRB light curves obtained by \textit{\textit{Swift}}/XRT between November 2004 and December 2010.
}
\maketitle{}
\addtocounter{page}{1} 
\section{Introduction}
The \textit{Swift} satellite \citep{swift}, launched on November 2004, during the past six years of operation detected and observed more than 600 Gamma-Ray Bursts (GRBs). The vast majority ($\sim$67\%) of these bursts were monitored in the soft X-ray band by the X-Ray Telescope (XRT, \citealt{ciao}) starting as early as $\sim$80 s after the trigger. The standard model explains the X-ray afterglow of GRBs as synchrotron radiation arising from the deceleration of a relativistic blast wave into the external medium.  The XRT follow up, therefore, is comprehensive of the tail of the prompt emission and of the afterglow. The XRT sample is now large enough to justify a statistical approach aimed at collecting, classifying and understanding the observational information of a wide and homogeneous sample of GRBs. This work will provide the most complete view of the X-ray properties of GRBs to date any existing theoretical model is asked to explain, while serving as a guide for future theoretical developments. The catalogue\footnote{Margutti et al. in preparation}  is now under completion, here we report the status of the work.

\section{Sample, Data Reduction and Analysis}
We analysed all the GRBs detected until the end of 2010, for which the afterglow had been observed by XRT with enough photons to extract a measurable spectrum. The sample consists of 437 GRBs out of a total 658 GRBs detected by \textit{Swift} of which 165 GRBs with redshift, 414 long GRBs (153 with z) and 23 short GRBs (12 with z). The original XRT data have been reduced with the method reported in \citet{margutti10}. We extracted the XRT light curves in the 0.3-10 keV XRT band as well as in the 0.3-1 keV, 1-2 keV, 2-3 keV and 3-10 keV band. Our XRT archive contains light curves in count-rate, flux and luminosity (for the redshift subsample) and all the relevant parameters of interest. The light curves in flux units are calibrated accounting for spectral evolution and the spectra have been derived using the NH column density estimated in a time interval where no spectral evolution is apparent.  
\section{Data analysis}
We fitted the light curves in flux units, using four functions:
\begin{itemize}
\item Single power-law:
\begin{eqnarray}
F(t)=N\,t^{-\alpha}.\label{plaw}
\end{eqnarray}
\item Smoothed broken power-law:
\begin{eqnarray}
F(t)=N\left( \left( \frac{t}{t_{b}}\right)^{-\frac{\alpha_1}{s}}+ \left(\frac{t}{t_{b}}\right)^{-\frac{\alpha_2}{s}} \right)^{s}.\label{bro1}
\end{eqnarray}
\item Sum of power-law and smoothed broken power-law:
\begin{eqnarray}
F(t)=N_1\,t^{-\alpha_1}+N_2\left( \left( \frac{t}{t_{b}}\right)^{-\frac{\alpha_2}{s}}+ \left(\frac{t}{t_{b}}\right)^{-\frac{\alpha_3}{s}} \right)^{s}.\label{bro4}
\end{eqnarray}
\item Sum of two smoothed broken power-laws:
\begin{eqnarray}
F(t)=N_1\left( \left( \frac{t}{t_{b1}}\right)^{-\frac{\alpha_1}{s_1}}+ \left(\frac{t}{t_{b1}}\right)^{-\frac{\alpha_2}{s_1}} \right)^{s_1}+N_2\left( \left( \frac{t}{t_{b3}}\right)^{-\frac{\alpha_3}{s_2}}+ \left(\frac{t}{t_{b3}}\right)^{-\frac{\alpha_4}{s_2}} \right)^{s_2}.\label{bro2}
\end{eqnarray}
\end{itemize}
where $\alpha$ is the decay power-law index, $t_b$ the break time, $s$ the smoothness parameter and $N$ the normalization. The best fit parameters were determined using the IDL Levenberg-Marquard least-squares fit routine (MPFIT) supplied by \citet{markwardt09}. All variability and fluctuations superimposed on the basic underlying light curve have been subtracted by iteration following \citep{margutti11}. Using the fit parameters, we calculated the total fluence (and the energy for the subsample of GRBs with known z) of our light curves as well as the one of different parts of the light curves ($E_1$, $E_2$, $E_3$, $E_4$), as shown in Fig. 1; moreover we calculate the fluence (energy when possible) of the excesses.

\section{Classification and morphology}
We classified the XRT light curves according to the fit function used, the presence of flares and the completeness of the curve (Fig. 2 and Tab. 1; see also \citealt{bernardini11}). For the different shapes (Fig. 1), we have: 
\begin{itemize}
\item \textbf{Type 0}: no breaks, Eq. \ref{plaw}; 
\item \textbf{Type Ia}: single break, Eq. \ref{bro1} and s $<$ 0; 
\item \textbf{Type Ib}: single break, Eq. \ref{bro1} and s $>$ 0; 
\item \textbf{Type IIa}: double broken power-law (steep-to-flat), Eq. \ref{bro4};
\item \textbf{Type IIb}: double broken power-law (flat-to-steep), Eq. \ref{bro4};
\item \textbf{Type III}: double broken power-law, Eq. \ref{bro2}.
\end{itemize}
A light curve is considered complete if the XRT re-pointing time is $<$300 s and the final count rate is comparable to the background. This ensures that the possible absence of the early steep decay is not due to an observational bias. See Tab. \ref{CLA_TA}  and Fig. 1, 2 for more details about our sample.

\begin{figure}  \label{shape}
 \begin{center}
    \includegraphics[width=1\textwidth]{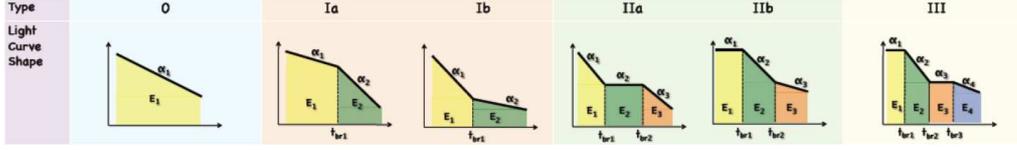}
  \caption{\small{Cartoon of the fit function used to classify and analyse the sample of XRT light curves. We also indicated some of the parameters derived and the time intervals for which we measured the emitted energy.}}
 \end{center}
\end{figure}

\begin{figure}  \label{classi}
  \begin{center}
    \includegraphics[width=0.6\textwidth]{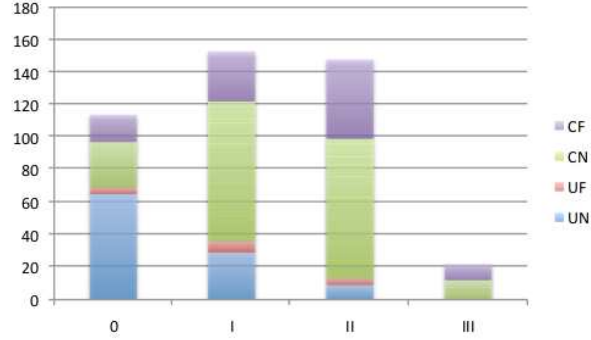}
     \caption{\small{Classification of the light curves in the sample, as illustrated in the Tab. \ref{CLA_TA}. 0, I, II, III indicate the fit function used where the number refers to the numbers of observed breaks in the light curve; for instance Type 0 means that the light curve is a pure power law. ``C'' stands for complete and ``U'' for incomplete; ``F'' indicate the presence of flares; on the contrary, there are not flares if in the name is present a ``N''. For example, ``ICN'' means that the fitting function used was a smoothed broken power law with a complete set of the data and no flares.}}
  \end{center}
\end{figure}

\begin{table}[h!]
\begin{center}
\begin{tabular}{c|cccc|c}
\hline
\hline
Type & UN & UF & CN& CF & Total \\
\hline
0 	& 65   & 4    & 28  & 17 & 114\\
I        & 29   & 7    & 86  &  31& 153\\
II       & 9     & 4    & 84  & 51 & 148\\
III      & 0     & 0    & 12  & 10  & 22\\      
\hline
Total&103 &15 & 212  & 107 & 437\\  
\hline
\hline
\end{tabular}
\caption{\small{Distribution of the GRBs light curves in the different classes. For more details see Fig. 2.}}\label{CLA_TA}
\end{center}
\end{table}%
 
 \section{Conclusion}
The temporal and spectral properties of a homogenous sample of 437 GRBs have been extracted. For the first time, we have a large and homogeneous data set that allows us to perform a statistical study of the X-ray properties of GRB afterglow. Notably, the sample of GRBs with redshift information comprises 165 elements: for this sub-sample it is possible to study the intrinsic properties of these explosions. \\ 
From Fig. 2 and Tab. 1, we notice that the probability to observe a Type I or Type II XRT light curve is higher if the data are complete (117/153 for Type I and 135/148 for Type II) evidencing an observational bias. The majority of incomplete light curves are Type 0 (69/118). In addition, almost all the incomplete light curves have not flares (103/118) and 33\% of the complete light curves have flares (107/319; see \citealt{chincarini10}).\\
As an example of the output of our analysis we plot the Dainotti relation \citep{dainotti08,dainotti10,dainotti11}, that is the relation between the end plateau luminosity and time, as we derive it from the row output and the distribution (histogram) of the total X-ray energy where the uncertainties for each bin have been computed by Monte Carlo simulations (Fig. 3).

\begin{figure}
  \begin{center}
    \includegraphics[width=0.4\textwidth]{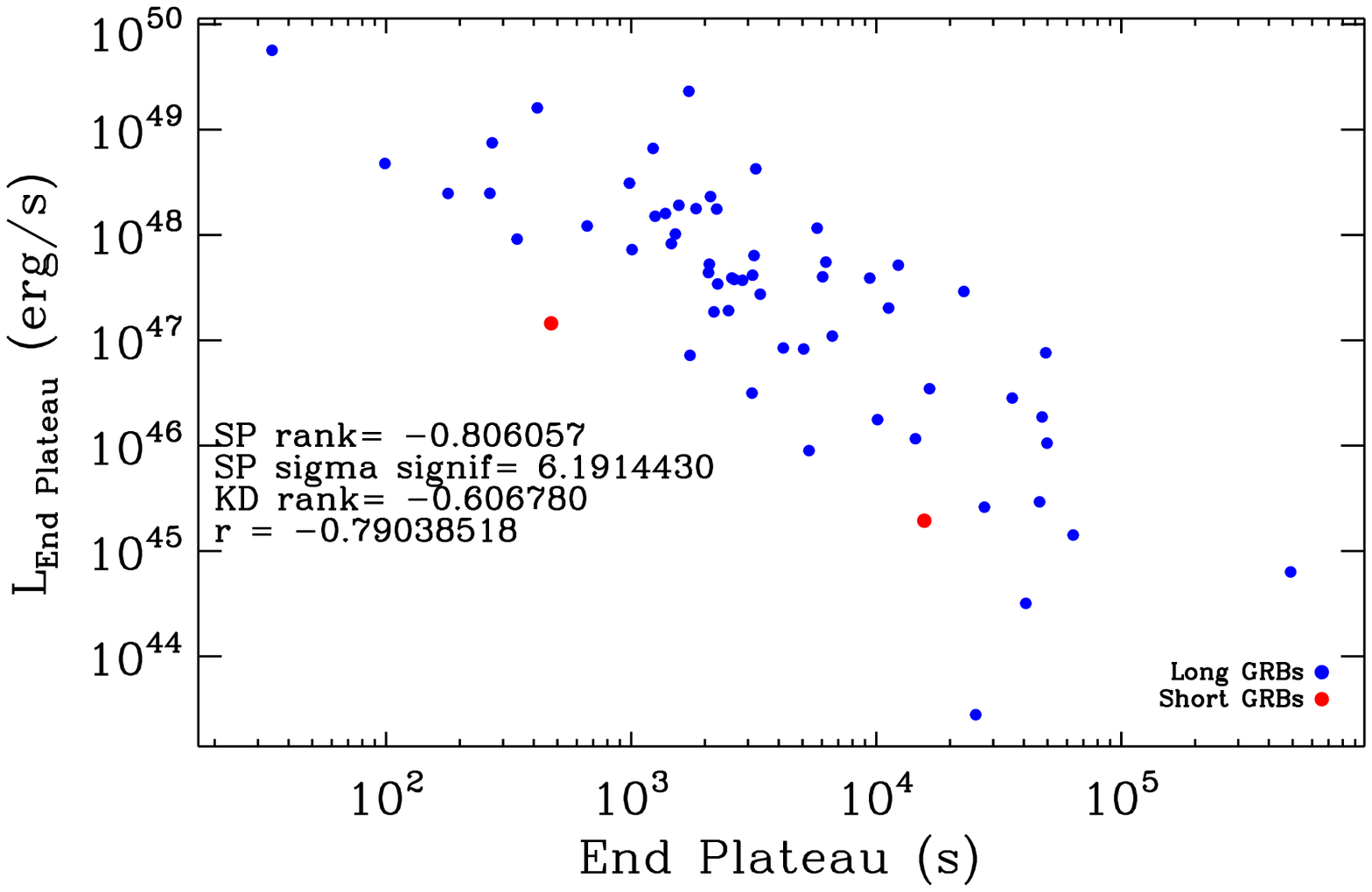}
    \includegraphics[width=0.4\textwidth]{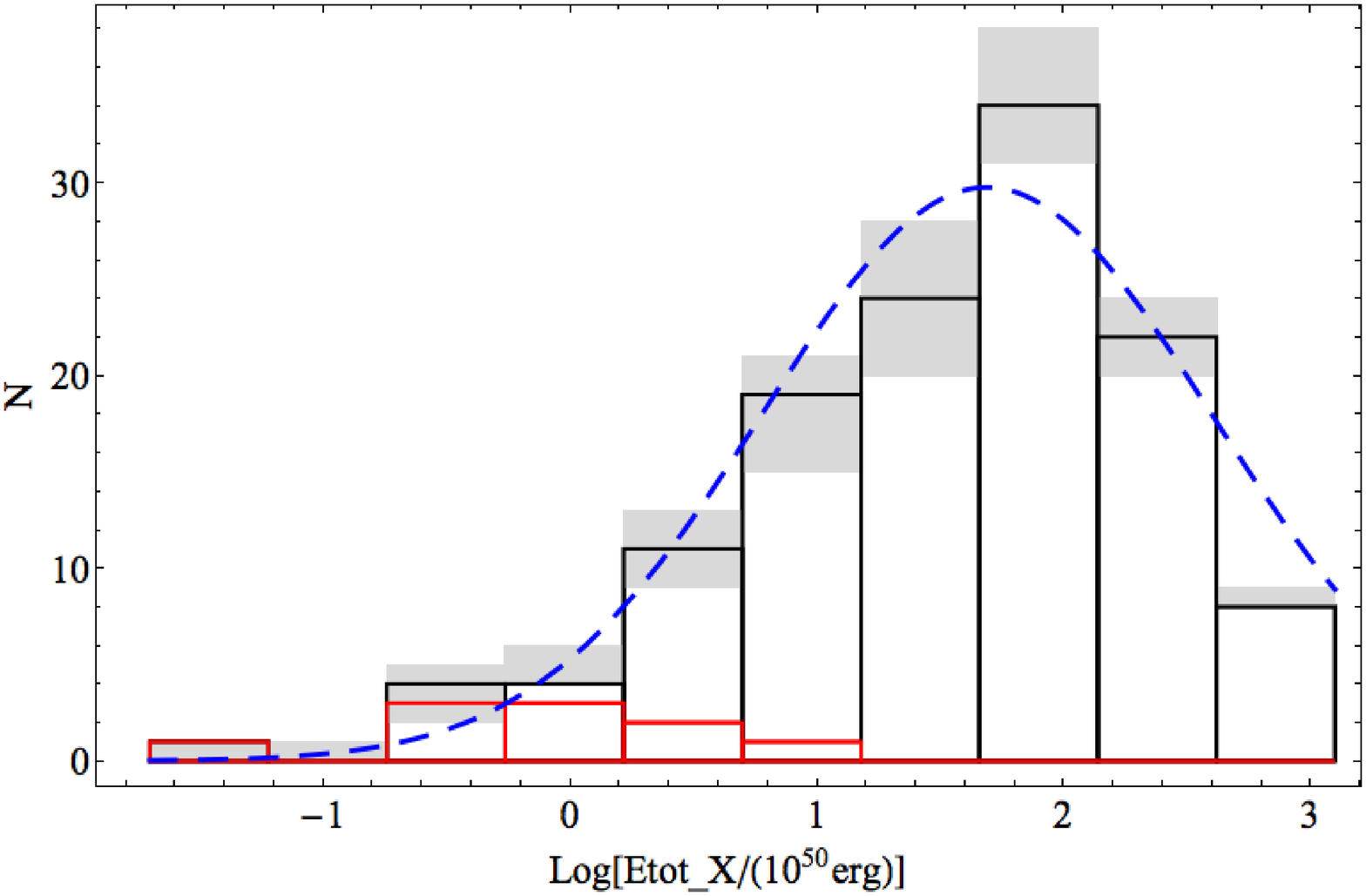}
     \caption{\small{The Dainotti relation (left) and the histrogram of the total X-ray energy with errors calculated with the Monte Carlo method (right). In red the short GRBs.}}
  \end{center}
  \label{esempi}
\end{figure}

\small{
\bibliographystyle{aa} 

}
\end{document}